\documentclass[10pt]{article}
\usepackage{amssymb}
\usepackage{graphicx}
\begin{document}
\newtheorem{prop}{}[section]
\newtheorem{defi}[prop]{}
\newtheorem{lemma}[prop]{}
\newtheorem{rema}[prop]{}
\newcommand{\boma}[1]{{\mbox{\boldmath $#1$}}}
\newcommand{\co}[2]{ {(#1)_{#2} \over #2!} }
\newcommand{\cp}[2]{ {(#1)_{#2} \over (#2)!} }
\def\sq{\bullet}
\def\sign{\mbox{sign}}
\def\dek{(-\delta)_k}
\def\defak{{(-\delta)_k \over k!}}
\def\den{(-\delta)_n}
\def\defan{{(-\delta)_n \over n!}}
\def\defun{{(1-\delta)_{n-1} \over (n-1)!}}
\def\v{{\bf{v}}}
\def\ffi{\phi}
\def\dfi{\varphi}
\def\Tau{{\mbox{\textsf{T}}} \hskip -0.08cm}
\def\ES{{\mathscr S}}
\def\J{{\mathscr J}}
\def\H{{\mathscr H}}
\def\K{{\mathscr K}}
\def\Kp{{\mathscr K}'}
\def\kp{k'}
\def\scrscr{\scriptscriptstyle}
\def\scr{\scriptstyle}
\def\dd{\displaystyle}
\def\z{\overline{z}}
\def\B{ B_{\mbox{\scriptsize{\textbf{C}}}} }
\def\Bc{ \overline{B}_{\mbox{\scriptsize{\textbf{C}}}} }
\def\ppartial{\overline{\partial}}
\def\d{\hat{d}}
\def\TT{T}
\def\G{ {\textbf G} }
\def\Hinf{ H^{\infty}(\reali^d, \complessi) }
\def\Hn{ H^{n}(\reali^d, \complessi) }
\def\Hm{ H^{m}(\reali^d, \complessi) }
\def\Ha{ H^{\d}(\reali^d, \complessi) }
\def\Ld{L^{2}(\reali^d, \complessi)}
\def\Lpi{L^{p}(\reali^d, \complessi)}
\def\Lq{L^{q}(\reali^d, \complessi)}
\def\Lr{L^{r}(\reali^d, \complessi)}
\def\Knb{K^{best}_n}
\def\k{\mbox{{\tt k}}}
\def\x{\mbox{{\tt x}}}
\def\D{\mbox{{\tt D}}}
\def\g{ {\textbf g} }
\def\QQQ{ {\textbf Q} }
\def\AAA{ {\textbf A} }
\def\gr{\mbox{graph}~}
\def\Q{$\mbox{Q}_a$~}
\def\PZ{$\mbox{P}^{0}_a$~}
\def\PZAL{$\mbox{P}^{0}_\alpha$~}
\def\PL{$\mbox{P}^{1/2}_a$~}
\def\PU{$\mbox{P}^{1}_a$~}
\def\PK{$\mbox{P}^{k}_a$~}
\def\PKU{$\mbox{P}^{k+1}_a$~}
\def\PI{$\mbox{P}^{i}_a$~}
\def\Pell{$\mbox{P}^{\ell}_a$~}
\def\PTM{$\mbox{P}^{3/2}_a$~}
\def\AZ{$\mbox{A}^{0}_r$~}
\def\AU{$\mbox{A}^{1}$~}
\def\lgraffa{ \mbox{\Large $\{$ } \hskip -0.2cm}
\def\rgraffa{ \mbox{\Large $\}$ } }
\def\restriction{ \stackrel{\setminus}{~}\!\!\!\!|~}
\def\M{{\scriptscriptstyle{M}}}
\def\m{m}
\def\Fre{Fr\'echet~}
\def\I{{\mathcal N}}
\def\ap{{\scriptscriptstyle{ap}}}
\def\fiap{\varphi_{\ap}}
\def\BBB{ {\textbf B} }
\def\EEE{ {\textbf E} }
\def\FFF{ {\textbf F} }
\def\TTT{ {\textbf T} }
\def\KKK{ {\textbf K} }
\def\FFi{ {\bf \Phi} }
\def\GGam{ {\bf \Gamma} }
\def\a{a}
\def\parn{\par\noindent}
\def\teta{M}
\def\elle{L}
\def\ro{\rho}
\def\al{\alpha}
\def\si{\sigma}
\def\be{\beta}
\def\ga{\gamma}
\def\de{\delta}
\def\la{\lambda}
\def\te{\vartheta}
\def\ch{\chi}
\def\et{\eta}
\def\complessi{{\textbf C}}
\def\reali{{\textbf R}}
\def\interi{{\textbf Z}}
\def\naturali{{\textbf N}}
\def\bT{{\textbf T}}
\def\T1{{\textbf T}^{1}}
\def\EE{{\mathcal E}}
\def\FF{{\mathcal F}}
\def\GG{{\mathcal G}}
\def\PP{{\mathcal P}}
\def\QQ{{\mathcal Q}}
\def\Np{{\hat{N}}}
\def\Lp{{\hat{L}}}
\def\Jp{{\hat{J}}}
\def\Pp{{\hat{P}}}
\def\Pip{{\hat{\Pi}}}
\def\Vp{{\hat{V}}}
\def\Ep{{\hat{E}}}
\def\Fp{{\hat{F}}}
\def\Gp{{\hat{G}}}
\def\Ip{{\hat{I}}}
\def\Tp{{\hat{T}}}
\def\Mp{{\hat{M}}}
\def\La{\Lambda}
\def\Ga{\Gamma}
\def\Si{\Sigma}
\def\Upsi{\Upsilon}
\def\Gag{{\check{\Gamma}}}
\def\Lap{{\hat{\Lambda}}}
\def\Sip{{\hat{\Sigma}}}
\def\Upsig{{\check{\Upsilon}}}
\def\Kg{{\check{K}}}
\def\ellp{{\hat{\ell}}}
\def\j{j}
\def\jp{{\hat{j}}}
\def\Stir{{\mathscr S}}
\def\BB{{\mathcal B}}
\def\LL{{\mathcal L}}
\def\SS{{\mathcal S}}
\def\DD{{\mathcal D}}
\def\VV{{\mathcal V}}
\def\WW{{\mathcal W}}
\def\OO{{\mathcal O}}
\def\CC{{\mathcal C}}
\def\RR{{\mathcal R}}
\def\AA{{\mathcal A}}
\def\CC{{\mathcal C}}
\def\JJ{{\mathcal J}}
\def\NN{{\mathcal N}}
\def\WW{{\mathcal W}}
\def\HH{{\mathcal H}}
\def\XX{{\mathcal X}}
\def\YY{{\mathcal Y}}
\def\ZZ{{\mathcal Z}}
\def\UU{{\mathcal U}}
\def\CC{{\mathcal C}}
\def\XX{{\mathcal X}}
\def\RR{{\mathcal R}}
\def\cir{{\scriptscriptstyle \circ}}
\def\circa{\thickapprox}
\def\vain{\rightarrow}
\def\ss{s}
\def\vains{\stackrel{\ss}{\rightarrow}}
\def\parn{\par \noindent}
\def\salto{\vskip 0.2truecm \noindent}
\def\spazio{\vskip 0.5truecm \noindent}
\def\vs1{\vskip 1cm \noindent}
\def\fine{\hfill $\diamond$ \vskip 0.2cm \noindent}
\newcommand{\rref}[1]{(\ref{#1})}
\def\beq{\begin{equation}}
\def\feq{\end{equation}}
\def\beqq{\begin{eqnarray}}
\def\feqq{\end{eqnarray}}
\def\barray{\begin{array}}
\def\farray{\end{array}}
\makeatletter
\@addtoreset{equation}{section}
\renewcommand{\theequation}{\thesection.\arabic{equation}}
\makeatother
\begin{center}
{\huge On the expansion of the Kummer function in terms of incomplete Gamma functions.}
\end{center}
\vspace{1truecm}
\begin{center}
{\large
Carlo Morosi${}^1$, Livio Pizzocchero${}^2$} \\
\vspace{0.5truecm}
${}^1$ Dipartimento di Matematica, Politecnico di
Milano, \\ P.za L. da Vinci 32, I-20133 Milano, Italy \\
e--mail: carmor@mate.polimi.it \\
${}^2$ Dipartimento di Matematica, Universit\`a di Milano\\
Via C. Saldini 50, I-20133 Milano, Italy\\
and Istituto Nazionale di Fisica Nucleare, Sezione di Milano, Italy \\
e--mail: livio.pizzocchero@mat.unimi.it
\end{center}
\vspace{1truecm}
\begin{abstract} The expansion of Kummer's hypergeometric
function as a series of incomplete Gamma functions is discussed,
for real values of the parameters and of the variable. The error performed
approximating the Kummer function with a finite sum of Gammas is evaluated
analytically. Bounds for it are derived, both pointwisely and uniformly in
the variable; these characterize the convergence rate of the series,
both pointwisely and in appropriate sup norms.
The same analysis shows that finite sums of very few Gammas
are sufficiently close to the Kummer
function.
The combination of these results with the known approximation methods
for the incomplete Gammas allows to construct upper and lower
approximants for the Kummer function using only exponentials, real powers  and rational functions.
Illustrative examples are provided.
\end{abstract}
\vspace{1truecm} \noindent \textbf{Keywords:} Confluent hypergeometric function, approximations and
expansions, inequalities in approximation.
\par \vspace{0.05truecm} \noindent \textbf{AMS 2000
Subject classifications:} 33C15, 41AXX.
\par \vspace{0.1truecm} \noindent \textbf{To appear in "Archives of Inequalities and Applications".}
\par
\section{Introduction and preliminaries.}
\label{intro}
Kummer's confluent hypergeometric function
$M(\alpha,\beta,x) \equiv {~}_{1} F_{1}(\alpha, \beta, x)$ can be introduced
in a number of equivalent ways \cite{Abr} .
For real values $\beta>\alpha>0$ of the parameters and for all real
$x$, we can regard this function to be defined by
\beq M(\alpha,\beta, x) := {1 \over B(\alpha, \beta-\alpha)}
\int_{0}^{1} d t~t^{\alpha-1} (1-t)^{\beta-\alpha-1} e^{x t}~,  \label{defint} \feq
where $B$ is the Beta function (see the notes following this Introduction).
The object of this paper is a series expansion for $M$ derived from the above integral
representation; for dealing more efficiently with this expansion, we will work on the
"reparametrized Kummer function"
\beq N(\alpha, \delta, x) := \int_{0}^1 d t~ t^{\alpha -1} (1 - t)^{\delta} e^{-x t}
\qquad \mbox{for $\alpha \in (0,+\infty)~, \delta \in (-1,+\infty),
x \in [0,+\infty)$}~. \label{repar} \feq
The knowledge of this function, with the variable $x$ confined to the
nonnegative semiaxis, allows to reconstruct the standard function $M$
on the full real axis: in fact,
for all $\alpha \in (0,+\infty)$,
$\beta \in (\alpha,+\infty)$ and $x \in [0,+\infty)$ it is
\beq M(\alpha, \beta, - x) = {1 \over B(\alpha, \beta-\alpha)}~
N(\alpha, \beta - \alpha - 1, x) \quad \mbox{and}~\feq
$$ M(\alpha, \beta, x) =
{1 \over B(\alpha, \beta-\alpha)}~
e^{x} N(\beta-\alpha, \alpha-1,x)~, $$
where the first equality follows immediately from the definition of $N$,
and the second one is inferred from the known identity \cite{Abr}
$M(\alpha, \beta, x) = e^{x} M(\beta-\alpha,\beta,-x)$. \parn
Let us fix the attention on $N$. If we expand in Taylor series
the term $(1-t)^{\delta}$ in the integral representation \rref{repar}
and integrate term by term, we obtain
\beq N(\alpha, \delta, x) =
\sum_{k=0}^{+\infty} \co{-\delta}{k} {\gamma(\alpha+k,x) \over x^{\alpha+k}}~,
\label{series} \feq
where the $\gamma$'s in the r.h.s. are incomplete Gamma functions, and
$(~)_k$ is the Pochhammer symbol (see again the notes at the end of the Section;
each summand in the above expansion also makes sense
for $x=0$, for it admits a finite $x \vain 0^{+}$ limit). A very simple case
of Eq. \rref{series} occurs when $\delta$ is integer: then $(1 - t)^{\delta}$
is a polynomial of degree $\delta$, and the above series for $N$ is a finite
sum with $k =0,1,..., \delta$. Also, it should be noted that for integer $\alpha$
all the above $\gamma$'s are elementary functions, expressible
in terms of exponentials and powers of $x$.
\parn
For large $x$, each incomplete function $\gamma(\alpha+k,x)$ approaches
the "complete" $\Gamma(\alpha+k)$; thus, one obtains a representation
\beq N(\alpha,\delta, x) \thickapprox
\sum_{k=0}^{+\infty} \co{-\delta}{k} {\Gamma(\alpha+k) \over x^{\alpha+k}}
\qquad \mbox{for $x \vain + \infty$}~, \label{poincar} \feq
which has been extensively studied \cite{Sla}; this must be dealt with carefully,
because (for noninteger $\delta$) the series in the r.h.s. is not convergent, but only asymptotic in the
Poincar\'e sense \cite{Olv}. \parn
In comparison with \rref{poincar}, the series \rref{series} seems to have attracted
less attention in the literature: for example, a recent survey \cite{Mul} on
computational methods for the Kummer function presents
\rref{series} as a "new method" for calculating this function.
An argument of \cite{Mul} proves convergence of the series for all
$x \in [0,+\infty)$; we think that it is worth to continue the analysis
of \rref{series} deriving accurate estimates on the error for
the series truncated at any finite order, and its rate of convergence.
\parn
Obtaining these estimates is the aim of the present paper. Our approach will
be based on some general inequalities about integrals of the form
$\int_{0}^b d t~ \psi(t) e^{-x t}$, of which Eq.\rref{repar} for $N$
gives an example. The behaviour of these integrals for $x \vain +\infty$ is
the object of the classical Watson Lemma \cite{Olv}; here we present a variant
of Watson's analysis, allowing to derive
upper and lower bounds on these integrals for all $x \geq 0$ by finite sums
of incomplete Gammas. These bounds are almost self-evident, but in fact
very useful in connection with $N$: after some technicalities, their application to
$N$ will yield the conclusions listed below. \parn
i) Upper and lower approximants of arbitrary accuracy can be derived for $N$, using
finite sums of incomplete Gammas. \parn
ii) The results i) imply convergence of the series \rref{series} on the
whole interval $[0,+\infty)$, not only pointwisely but also uniformly in $x$;
uniform convergence can be expressed using appropriate sup-norms. \parn
iii) The approximants for $N$ obtained in i) are accurate on the
whole semiaxis $[0,+\infty)$ even when one sums very few terms (however,
for very small $x$ the Taylor expansion of $N$ is generally more precise). \parn
iv) The accuracy of the low order expansions in incomplete Gammas is essentially preserved if,
for noninteger $\alpha$, one replaces these functions with some known approximants of
Pad\'e or other types. In conclusion, it is possible to obtain accurate upper and lower approximants
for $N$ for both integer and noninteger $\alpha$
using only exponentials, real powers of $x$ and rational functions. \parn
Let us outline the organization of the paper. The rest of this Section fixes some notations
and reviews some basic facts about Gamma functions and their approximations, the Beta
function and the Pochhammer
symbols employed extensively in the sequel. Sect.\ref{main} presents the upper and
lower bounds for a general integral of the Watson type in terms of incomplete Gammas.
Sect.\ref{remind} analyses the error in the Taylor expansion of the funtion
$t \mapsto (1-t)^{\delta}$, globally on the interval $[0,1)$; this is necessary
for applying the bounds of Sect.\ref{main} to the integral representation
of $N$. Sect.\ref{kum} contains the main results about the expansion of $N$ by
incomplete Gammas, and the error for the expansion truncated at any finite order.
Bounds for the error are derived analytically,
both pointwisely and uniformly in $x$; some numerical tests on these theoretical bounds are presented,
for an appreciation of their reliability.
Sect.\ref{exa} illustrates by some examples the low order
expansions for $N$, and the elementary substitutes derived from them replacing the incomplete Gammas
with their known approximants by exponentials, powers of $x$ and rational functions.
\vskip 0.2cm \noindent
\textbf{Miscellaneous notations.} Throughout the paper $\naturali$
and $\naturali_0$ are, respectively, the nonnegative and
the positive integers.  The symbol $\sim$ indicates
that two functions or sequences are asymptotic in the elementary sense,
i.e., that their ratio tends to unity.
\parn
\textbf{On Gamma functions.} We use
the standard notations for the incomplete
and "complete" Gamma functions; in the ranges indicated below, we can
regard them to be defined by
\beq \gamma(\nu, x) := \int_{0}^{x} d s~s^{\nu -1} e^{-s}, \quad
\Gamma(\nu) := \int_{0}^{+\infty} d s~s^{\nu -1} e^{-s} \qquad
\mbox{for $\nu \in (0,+\infty), x \in [0,+\infty)$}. \label{equa} \feq
For either $\nu$ or $x$ fixed, the incomplete function has the asymptotics
\beq \gamma(\nu, x)~ \left\{ \barray{lll} \sim x^{\nu} /\nu
&& \mbox{for $x \vain 0^{+}$,} \\ \vain \Gamma(\nu) && \mbox{for $x \vain +\infty$~;}
\farray \right. \label{asi} \feq
\beq \gamma(\nu, x) \sim {x^{\nu} \over \nu} e^{-x}
\qquad \mbox{for $\nu \vain + \infty$}~. \label{asial} \feq
Of course,
\beq {d \over d x} \gamma(\nu,x) = x^{\nu-1} e^{-x}~; \label{deriga} \feq
also, we have the recursion rule
\beq \gamma(\nu, x) = (\nu-1) \gamma(\nu-1, x) - x^{\nu -1} e^{-x}~.
\label{regam} \feq
Concerning the complete function, we will continuously use the relations
\beq \Gamma(\nu+1) = \nu \Gamma(\nu) \label{rree} \feq
and $\Gamma(n+1) = n!$ for $n \in \naturali$.
Sometimes we need the standard extension of the complete function $\Gamma$
from $[0,+\infty)$ to $\reali \setminus (- \naturali)$, which is uniquely
determined asking \rref{rree} to hold everywhere on this enlarged domain. It is
$\Gamma(\nu) \vain \infty$ for $\nu \vain p \in -\naturali$.
\parn
\textbf{Practical evaluation of $\boma{\gamma(\nu,x)}$ for fixed
$\boma{\nu \in (0,+\infty)}$}. This is very simple for integer $\nu$: using
the recursion rule \rref{regam}, we reduce the computation to the case
\beq \gamma(1, x) = 1 - e^{-x}~. \feq
For arbitrary $\nu$ one constructs approximants of $\gamma(\nu,x)$ of different types, for
$x$ small or large. Some elementary approximants are derived from Taylor's formula
applied to the exponential $e^{-s}$ in Eq.\rref{equa}. For $s \geq 0$ and $m \in \naturali_0$,
we have $e^{-s} = \sum_{j=0}^{m-1} (-1)^j s^{j}/j! + (-1)^m e^{-c} s^m/m!$, where
$c \in [0,s]$; of course, $e^{-x} \leq e^{-c} \leq 1$ if $s \in [0,x]$. Thus
\beq \sum_{j=0}^{m-1} {(-1)^j \over j!} s^{j} +
{u_m(x) \over m!} s^m \leq e^{-s} \leq \sum_{j=0}^{m-1} {(-1)^j \over j!} s^{j} + {U_m(x) \over m!} s^m
\label{ems} \feq
for $x \in [0,+\infty)$, $s \in [0,x]$, $m \in \naturali_0$, where
\beq u_m(x) := \min \Big( (-1)^m, (-1)^m e^{-x} \Big) = \left\{ \barray{lll} e^{-x} && \mbox{for $m$ even,} \\
1 && \mbox{for $m$ odd;} \farray \right. \label{up} \feq
\beq U_m(x) := \max \Big( (-1)^m, (-1)^m e^{-x} \Big) = \left\{ \barray{lll} 1 && \mbox{for $m$ even,} \\
- e^{-x} && \mbox{for $m$ odd;} \farray \right. \label{uup} \feq
We substitute the inequalities \rref{ems} in the definition \rref{equa} for $\gamma$, and
integrate term by term. This gives
\beq \tau_m(\nu,x) \leq \gamma(\nu,x) \leq \Tau_m(\nu,x) \qquad \mbox{for
$\nu \in (0,+\infty)$, $x \in [0,+\infty)$, $m \in \naturali_0$}~, \label{tay1} \feq
\beq \tau_{m}(\nu,x) := \sum_{j=0}^{m-1} {(-1)^j \over j!} {x^{\nu+j} \over \nu + j} +
{u_m(x) \over m!} {x^{\nu+m} \over \nu + m},~~\label{tay2} \feq
$$ \Tau_{m}(\nu,x) := \sum_{j=0}^{m-1} {(-1)^j \over j!} {x^{\nu+j} \over \nu + j} +
{U_m(x) \over m!} {x^{\nu+m} \over \nu + m}~. $$
The "Taylor" approximants $\tau_m$, $\Tau_m$ are interesting for small $x$ only; for large $x$, a
completely different approach must be employed. As a preliminary, one observes that
repeated application of the recursion rule \rref{regam} starting from a noninteger value
of $\nu$ reduces the problem to the case $\nu \in (0,1)$: in this case
reliable approximants are known for large $x$, mainly of two types. \parn
Firstly, there is a sequence of "Laurent" approximants,
involving negative powers of $x$. These are constructed working on the representation
$\gamma(\nu,x)= \Gamma(\nu) - \int_{x}^{+\infty} d s~s^{\nu-1} e^{-s}$; by
repeated integrations by parts, one infers \cite{Olv}
\beq \Gamma(\nu) - x^{\nu-1} e^{-x} \Lambda_q(\nu,x) \leq \gamma(\nu,x)
\leq \Gamma(\nu) - x^{\nu-1} e^{-x} \lambda_q(\nu,x) \feq
for all $\nu \in (0,1)$, $x \in (0,+\infty)$ and $q \in \naturali_0$, where
\beq \lambda_q(\nu, x) := \sum_{k=0}^{q-1} {\langle \nu-1 \rangle_k \over x^k} +
{\langle \nu-1 \rangle^{-}_q \over x^q},
\quad \Lambda_q(\nu, x) := \sum_{k=0}^{q-1} {\langle \nu-1 \rangle_k \over x^k} +
{\langle \nu-1 \rangle^{+}_q \over x^q}~; \feq
\beq
\langle \mu \rangle_0 := 1~, \qquad \langle \mu \rangle _h :=
\mu (\mu-1) (\mu-2)... (\mu-h+1)~~~\mbox{for $h \in \naturali_0$},~\feq
$$ \chi^{+} := \max(\chi,0)~, \qquad \chi^{-} := \min(\chi,0) \qquad \mbox{for $\chi \in \reali$}. $$
Typically, these Laurent approximants are accurate for very large $x$ but less satisfactory
for intermediate values, close to $1$. This defect is overcome by the Pad\'e approximants,
constructed
by more refined techniques \cite{Luk}: these have the form
\beq \Gamma(\nu) - x^{\nu-1} e^{-x} \Pi_q(\nu,x) \leq \gamma(\nu,x)
\leq \Gamma(\nu) - x^{\nu-1} e^{-x} \pi_q(\nu,x) \label{pad1} \feq
for $\nu \in (0,1)$, $x \in (0,+\infty)$ and $q \in \naturali$, where both $\Pi_q$, $\pi_q$ are
ratios of polynomials of degree $q$ in $x$; the lowest order cases are
\beq \pi_0(\nu,x) := 0,~\Pi_0(\nu,x) := 1~; \quad
\pi_1(\nu,x) := {x \over x + 1 - \nu},~\Pi_1(\nu,x) :=
{x +1 \over x + 2 - \nu}~; \label{pad2} \feq
$$ \pi_2(\nu,x) := {x (x + 3 - \nu) \over x^2 + 2 (2 - \nu) x + (1 - \nu)(2 - \nu)},
\quad \Pi_2(\nu,x) := {x^2 + (5 - \nu) x + 2 \over x^2 + 2 (3 - \nu) x + (2 - \nu)(3 - \nu)}~. $$
Of course, one can match the approximants of the above different types
to get globally accurate estimates. For example, let us evaluate $\gamma(1/2,x)$
using the Taylor bounds of order
$m=4$, and the Pad\'e bounds of order $q=1$; this
gives the inequalities $h(x) \leq \gamma(1/2,x) \leq H(x)$ where $h$ is the maximum of
the two lower bounds, and $H$ the minimum of the two upper bounds
($h$ is the Taylor lower bound for $0 \leq x \leq 1.16$,
and the Pad\'e one for $x \geq 1.17$; $H$ is the Taylor upper bound for
$0 \leq x \leq 1.48$, and the Pad\'e one for $x \geq 1.49$). The relative uncertainty
$(H-h)/(H+h)$ is $< 0.005$ for all $x \in [0,+\infty)$, and attains its maximum
at a point $x_0 \in (1.48,1.49)$.
\parn
\textbf{Pochhammer's symbol.} For $\alpha \in \reali$ and $k \in \naturali$, this is defined setting
\beq (\alpha)_0 := 1~~, \qquad (\alpha)_k := \alpha (\alpha+1) (\alpha+2) ... (\alpha+k-1)~; \feq
one has
\beq (\alpha)_{k+1} = (\alpha)_{k} (\alpha+k) = \alpha (\alpha+1)_{k}~, \qquad
{(\alpha+1)_m \over m!} = \sum_{k=0}^m {(\alpha)_k \over k!} \label{repoc} \feq
(the first two equalities are elementary, the last one is proved recursively).
It must be noted that
\beq (\alpha)_k = {\Gamma(\alpha+k) \over \Gamma(\alpha)}~, \label{pogam} \feq
(provided that, for $\alpha \in -\naturali$, the r.h.s. be intended as a limit
from noninteger values; incidentally, the above relation allows to compute
$\Gamma(1/2+k)$ starting from $\Gamma(1/2) = \sqrt{\pi}$).
The behaviour of Pochhammer's symbol for real $\alpha$ and large
$k \in \naturali$ is described by the equation
\beq (\alpha)_k~\left\{ \barray{lll} \sim k!~ k^{\alpha-1}/\Gamma(\alpha)
&& \mbox{for $\alpha \not\in -\naturali$, $k \vain +\infty~$,} \\
=0 && \mbox{for $\alpha \in -\naturali, k \geq 1 -\alpha$~;} \farray
\right.
\label{asipoc} \feq
(the statement for $\alpha \in -\naturali$ is evident, the other one follows from
known properties of the Gamma function \cite{Olv}). \parn
\textbf{The Beta function $\boma{B}$.} The standard definition reads \cite{Abr}
\beq B(\mu, \nu) := \int_{0}^1 d t~ t^{\mu-1} (1 - t)^{\nu-1} \qquad \mbox{for $\mu, \nu \in (0,+\infty)$};
\label{beta} \feq
this function is related to the complete Gamma by the identity
\beq B(\mu, \nu) = {\Gamma(\mu) \Gamma(\nu) \over \Gamma(\mu + \nu)}~. \label{ideb} \feq
\textbf{Elementary facts about $\boma{N}$.} Throughout the paper we stick to the
definition \rref{repar} of $N(\alpha, \delta, x)$, for all $\alpha \in (0,+\infty)$,
$\delta \in (-1,+\infty)$ and $x \in [0,+\infty)$. Even though our main interest is the
representation \rref{series} for $N$ in terms of incomplete Gammas, it is convenient
to review the Taylor expansion of $N$ that will be sometimes compared with \rref{series}.
The derivation of the Taylor expansion is elementary: we substitute in Eq.
\rref{repar} the inequalities \rref{ems}, with $s= t x$, and we integrate
term by term. Recalling the definition \rref{beta} of $B$, we obtain the final result
\beq t_m(\alpha, \delta, x) \leq N(\alpha, \delta, x) \leq T_m(\alpha, \delta, x) \label{tayn1} \feq
for $\alpha \in (0,+\infty)$, $\delta \in (-1,+\infty)$, $x \in [0,+\infty)$ and $m \in \naturali_0$,
where
\beq t_m(\alpha, \delta, x) := \sum_{k=0}^{m-1} {(-1)^k \over k!} B(\alpha + k, \delta + 1) x^k + {u_m(x) \over m!}
B(\alpha + m, \delta + 1) x^m~, \label{tayn2} \feq
$$ T_m(\alpha, \delta, x) := \sum_{k=0}^{m-1} {(-1)^k \over k!} B(\alpha + k, \delta + 1) x^k + {U_m(x) \over m!}
B(\alpha + m, \delta + 1) x^m,$$
and $u_m$, $U_m$ are as in Eq.s (\ref{up}-\ref{uup}).
Sending $m$ to $+\infty$ in Eq. \rref{tayn1}, we obtain the power series representation
\beq N(\alpha, \delta, x) = \sum_{k=0}^{+\infty} {(-1)^k \over k!} B(\alpha + k, \delta + 1) x^k~,
\label{serta} \feq
converging for all $x$ in the chosen domain $[0,+\infty)$. The power series expansion is
often used as a definition of the Kummer function, alternative to the integral
representation adopted here as our starting point. As a final remark about \rref{serta},
we note that Eq.s \rref{ideb} \rref{pogam} imply
\beq B(\alpha+k, \delta+1) = {(\alpha)_k \over (\alpha+\delta+1)_k} B(\alpha, \delta+1)~. \feq
\section{Integrals of the Watson type.}
\label{main}
By a Watson type integral, we mean an (improper) Riemann integral depending on a
real parameter $x$, of the form
\beq W(x) := \int_{0}^b d t~\psi(t)~ e^{-x t}~, \qquad
b \in (0,+\infty],~\psi \in C((0, b), \reali)~;
\label{int}  \feq
in the sequel, the expression "$W(x)$ exists" will mean that the integral
\rref{int} is convergent, for the considered value of the parameter $x$.
Watson's Lemma \cite{Olv} was originally formulated as a statement on
the behaviour of $W(x)$ for $x \vain +\infty$, under appropriate conditions
on $\psi$.
Here we present a simple Watson-type statement
giving upper or lower bounds on $W(x)$, \textsl{without sending $x$ to infinity}:
this is an \textsl{inequality}, holding for both small and large $x$.
\begin{prop}
\label{wat}
\textbf{Proposition.} Assume
\beq \psi(t) \geq (\mbox{resp.} \leq) \sum_{k=0}^n p_k t^{\lambda_k}
\qquad \forall t \in (0,b), \label{supp} \feq
where $n \in \naturali$, $p_k \in \reali$, $\lambda_k \in (-1, +\infty)$ for $k=1,...,n$.
If $x \in (0,+\infty)$ and $W(x)$ exists, then
\beq W(x) \geq (\mbox{resp.} \leq)~
\sum_{k=0}^n {p_k} {\gamma(\lambda_{k}+1, b x) \over x^{\lambda_{k}+1 } }~.
\label{maineq} \feq
In the above equation, for $b = +\infty$
one intends $b x := +\infty$, $\gamma(\lambda_{k}+1, +\infty) :=
\Gamma(\lambda_{k}+1)$. \parn
If $b < +\infty$ and $W(0)$ exists, the inequality \rref{maineq}
holds for it replacing each term $\gamma(\lambda_{k}+1, b x) / x^{\lambda_{k}+1 }$
with its $x \vain 0^{+}$ limit, which equals $b^{\lambda_k+1}/(\lambda_k+1)$.
\end{prop}
\textbf{Proof.} We simply insert the inequality \rref{supp} in the definition
\rref{int} of $W(x)$; this gives $W(x) \geq $ (resp. $\leq$) $\sum_{k=0}^n p_k
\int_{0}^b d t~ t^{\lambda_k} e^{-x t}$. For $x \in (0,+\infty)$ and
$b$ either finite or infinite,
the variable change $t = s/x$ in the integrals yields the thesis. \parn
For $x=0$, and $b$ finite, each integral in the previous sum equals
$b^{\lambda_k+1}/(\lambda_k+1)$, which is the $x \vain 0^{+}$ limit of
$\gamma(\lambda_{k}+1, b x) / x^{\lambda_{k}+1 } $ due to \rref{asi}.
\fine
\section{The error in Taylor's expansion of $\boma{(1-t)^{\delta}}$.}
\label{remind}
The function $N(\alpha, \beta, x)$ of Eq.\rref{repar}
is a Watson type integral of the form \rref{int}, with $b=1$ and
$\psi(t) \equiv \psi(\alpha,\delta,t) := t^{\alpha-1} (1 - t)^{\delta}$. \parn
We want to find both upper and lower bounds for $\psi$ in terms of powers of $t$, of the
form \rref{supp}; the problem can be reduced to studying the Taylor expansion about $t=0$ of the function
\beq t \in [0,1) \mapsto (1 - t)^{\delta} \qquad \in C^{\infty}([0,1), \reali)~. \feq
For our purposes, it is
essential to discuss carefully the behaviour of the error \textsl{everywhere on the interval} $t \in [0,1)$;
this is the object of the following
\begin{prop}
\label{tayp}
\textbf{Lemma.} For all $\delta \in \reali$ and
$n \in \naturali_0$, there is a unique
function $\rho_n(\delta,\sq) \in C^{\infty}([0,1), \reali)$ such that
\beq (1 - t)^{\delta} = \sum_{k=0}^{n-1} \co{-\delta}{k} t^k + \rho_n(\delta, t) t^n \qquad
\forall t \in [0,1)~. \label{tayeq} \feq
It is
\beq \rho_n(\delta, t) = {(-\delta)_n \over (n-1)!}~
\int_{0}^1 d u~ (1 -u)^{n-1}~(1 - t u)^{\delta -n}~, \quad
\sign~\rho_n(\delta, t) = \mbox{const.} = \sign~ (-\delta)_{n}~. \label{espron} \feq
Furthermore,
\beq \rho_{n}(\delta, 0) = \co{-\delta}{n}~; \quad
\lim_{t \vain 1^{-}} \rho_{n}(\delta, t) = \left\{ \barray{lll} - (1- \delta)_{n-1}/(n-1)!
&& \mbox{if~ $\delta>0$,} \\ 0 && \mbox{if~ $\delta=0$,}
\\ + \infty && \mbox{if~ $\delta <0$.}
\farray \right. \label{eqlim} \feq
For all $t \in [0,1)$, the derivative of $\rho_n(\delta, \sq)$ is described by
\beq \rho'_n(\delta, t) =
{(-\delta)_{n+1} \over (n-1)!}~
\int_{0}^1 d u~ u (1 -u)^{n-1}~(1 - t u)^{\delta -n-1}~, \label{derint} \feq
$$ \quad \sign~\rho'_n(\delta, t) = \mbox{const.} = \sign~ (-\delta)_{n+1}~. $$
(All statements on the signs include the zero case, intending $\sign(0) :=0$.)
\end{prop}
\textbf{Proof.} Taylor's formula for an arbitrary $f \in C^{m}([0,b),\reali)$,
($m \geq n$) ensures the
existence of a unique function $\rho_n \in C^{m-n}([0,b), \reali)$ such that
\beq f(t) = \sum_{k=0}^{n-1} {1 \over k!} f^{(k)}(0) t^k + \rho_n(t) t^n
\qquad \mbox{for $t \in [0,b)$}~, \label{taygen} \feq
which has the integral representation
\beq \rho_n(t) = {1 \over (n-1)!}~\int_{0}^{1} d u~(1-u)^{n-1}~ f^{(n)}(t u)~. \feq
The function $f(t) \equiv f(\delta, t) = (1-t)^{\delta}$ on $[0,1)$ has derivatives
\beq f^{(k)}(t) = {d^k \over d t^k}~(1 - t)^{\delta}  =
(-\delta)_{k} (1 - t)^{\delta - k}~; \label{der} \feq
this yields Eq.s \rref{tayeq}, \rref{espron}. The integral in Eq.\rref{espron} is clearly
positive, so $\rho_n(\delta, t)$ has the sign of the coefficient $(-\delta)_n$. \parn
In general, for the function $\rho_n$ of the expansion \rref{taygen} we have $\rho_n(0)= f^{(n)}(0)/n!$,
which equals $(-\delta)_n/n!$ in the present case. The results
\rref{eqlim} for $\lim_{t \vain 1^{-}} \rho_n(\delta,t)$ can be derived from \rref{tayeq}:
if $\delta >0$, the $t \vain 1^{-}$ limit computed
from \rref{tayeq} is in fact $-\sum_{k=0}^{n-1} (-\delta)_k/k!$, but this equals
$-(1-\delta)_{n-1}/(n-1)!$ due to the third relation \rref{repoc}.
(This case could be treated alternatively by the integral representation
\rref{espron}, which implies
$$ \lim_{t \vain 1^{-}} \rho_n(\delta, t) =
{(-\delta)_n \over (n-1)!}~\int_{0}^1 d u~ (1 -u)^{\delta -1} ={(-\delta)_n \over (n-1)!} {1 \over \delta}
= - {(1-\delta)_{n-1} \over (n-1)!} $$
due to the second relation \rref{repoc}). \parn
To obtain the representation \rref{derint} for $\rho'_{n}(\delta,t)$, one derivates
Eq.\rref{espron} and then employs the first identity \rref{repoc} to express the
coefficient before the integral in the form $(- \delta)_{n+1}/(n-1)!$. The integral
in \rref{derint} is positive, so the derivative $\rho'_n$ has the sign of $(- \delta)_{n+1}$.
\fine
\textbf{Remark.} $\rho_n(\delta, \sq)$ and its derivative
are hypergeometric functions of the Gaussian type \cite{Abr}. The Gaussian hypergeometric
function $F \equiv {~}_2 F_{1}$ with parameters $a,b,c$ can be defined setting
\beq F(a,b,c;t) :=
{1 \over B(b, c-b)}~\int_{0}^1 d u~u^{b-1}
(1 - u)^{c-b-1} (1 - t u)^{-a}~. \feq
By comparison, we see that
\beq \rho_n(\delta, t) = \co{-\delta}{n}~F(n-\delta,1,n+1;t)~, \qquad
\rho'_n(\delta, t) = \cp{-\delta}{n+1}~
F(n+1-\delta,2,n+2;t)~. \feq
\fine
The previous Lemma implies
\begin{prop}
\label{coro}
\textbf{Corollary.} For $\delta \in \reali$ and $n \in \naturali_0$, consider again the function
$\rho_n(\delta,\sq)$ of Eq.\rref{tayeq}; then, its $\sup$ and $\inf$ are as follows. \parn
For $\delta \geq 0$,
\beq \inf_{t \in [0,1)} \rho_{n}(\delta, t) = r_n(\delta), \quad
\sup_{t \in [0,1)} \rho_{n}(\delta, t) = R_n(\delta)~, \feq
\beq r_n(\delta) := \min \left( \co{-\delta}{n}, - \cp{1-\delta}{n-1} \right) =
\left\{ \barray{lll} \mbox{the left term} && \mbox{if~ $(-\delta)_{n+1} \geq 0$,} \\
\mbox{the right term} && \mbox{if~ $(-\delta)_{n+1} <0$,}
\farray \right. \label{rn} \feq
\beq R_n(\delta) := \max \left( \co{-\delta}{n}, - \cp{1-\delta}{n-1} \right) =
\left\{ \barray{lll} \mbox{the right term} && \mbox{if~ $(-\delta)_{n+1} > 0$,} \\
\mbox{the left term} && \mbox{if~ $(-\delta)_{n+1} \leq 0$.}
\farray \right. \label{rrnn} \feq
For $\delta <0$, it is
\beq \inf_{t \in [0,1)} \rho_{n}(\delta, t) = r_n(\delta)~, \qquad
\sup_{t \in [0,1)} \rho_{n}(\delta, t) = + \infty~, \feq
with
\beq r_n(\delta) := \co{-\delta}{n}~. \label{agree} \feq
\end{prop}
\textbf{Proof.} In any case $\rho_n(\delta, \sq)$ is monotonic, so its $\sup$ and $\inf$ are
the limit values at $t = 0$ or $t=1$. The thesis follows from the results of the previous
Lemma about these limits and the sign of $\rho'_n(\delta, \sq)$. \fine
\textbf{Remarks.} i) For all $n
\in \naturali_0$, it is
\beq - \cp{1-\delta}{n-1} = {n \over \delta} \co{-\delta}{n}~~\mbox{
if $\delta \in \reali \setminus \{ 0 \}$}~; \qquad - \cp{1-\delta}{n-1} = -1,~
\co{-\delta}{n} = 0~~\mbox{if $\delta = 0$} \label{eqrem} \feq
(the first relation above follows from the second identity \rref{repoc}).
\parn
ii) Many of the previous statements involve the signs of the Pochhammer symbols
of $-\delta$; therefore, the following tables can be useful. For a real noninteger $\delta$,
with integer part $[\delta]$, and for all $k \in \naturali$ it is
\beq \sign(-\delta)_k = \left\{ \barray{lll} (-1)^k
&& \mbox{if~ $\delta \geq 0$, $0 \leq k \leq [\delta]$,} \\
(-1)^{[\delta]+1} && \mbox{if~ $\delta \geq 0$, $k \geq [\delta] + 1$,}
\\ 1 && \mbox{if~ $\delta < 0$.}
\farray \right. \label{eqsign1} \feq
For $\delta$ a relative integer and $k \in \naturali$,
\beq \sign(-\delta)_k = \left\{ \barray{lll} (-1)^k
&& \mbox{if~ $\delta \geq 0$, $0 \leq k \leq \delta$,} \\ 0 && \mbox{if~ $\delta \geq 0$,
$k \geq \delta + 1$,}
\\ 1 && \mbox{if~ $\delta < 0$.}
\farray \right. \label{eqsign2} \feq
(Of course, the vanishing of $(-\delta)_k$ for $\delta \in \naturali$ and $k \geq \delta + 1$
reflects the fact that, in this case, $(1-t)^\delta$ is a polynomial of degree $\delta$). \fine
\section{Expanding the Kummer function.}
\label{kum}
\begin{figure}
\begin{center}
\includegraphics[
height=2.0in,
width=2.8in
]%
{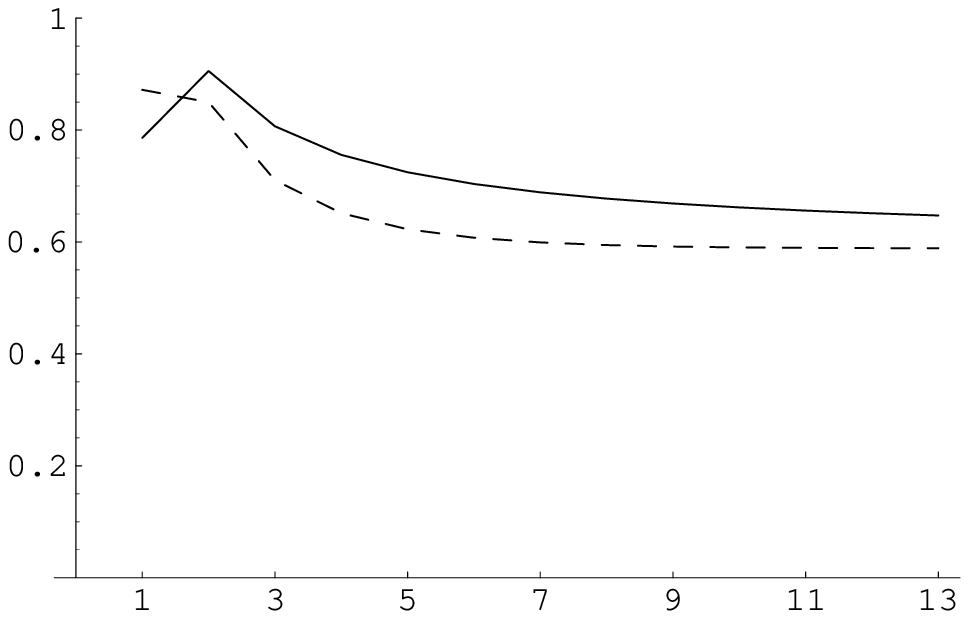}%
\caption{Ratios $| \epsilon_n |/\EE_n$ for $\alpha=2$, $\delta=3/2$, $x=1$ (continuous line) and
$x=6$ (dashed line).}
\label{g1}
\end{center}
\begin{center}
\includegraphics[
height=2.0in,
width=2.8in
]%
{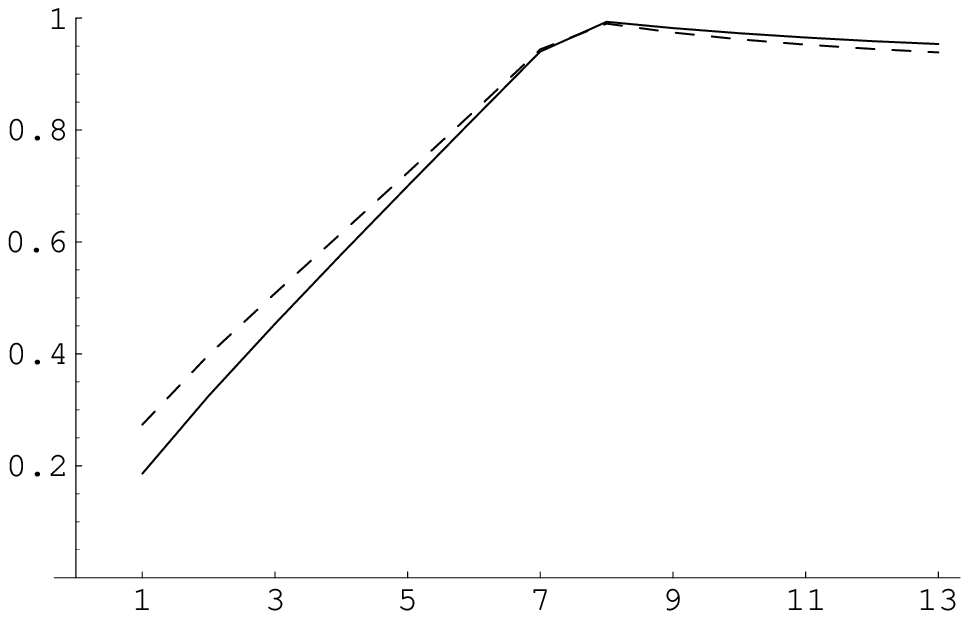}%
\caption{Ratios $| \epsilon_n |/\EE_n$ for $\alpha=3$, $\delta=15/2$, $x=1$ (continuous line) and
$x=6$ (dashed line).}
\label{g2}
\end{center}
\begin{center}
\includegraphics[
height=2.0in,
width=2.8in
]%
{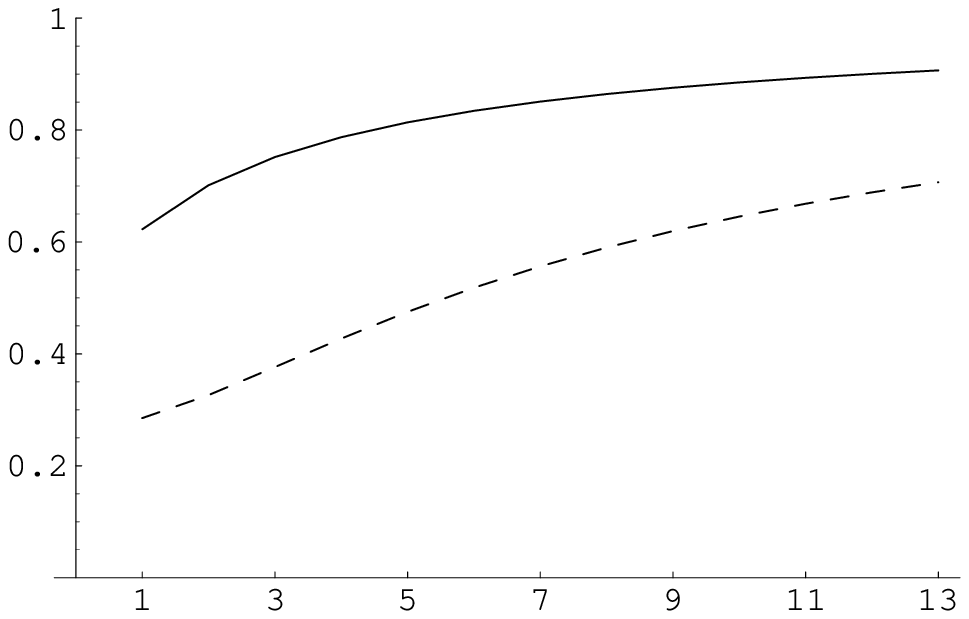}%
\caption{Ratios $| \epsilon_n |/\EE_n$ for $\alpha=2$, $\delta=-1/2$, $x=1$ (continuous line) and
$x=6$ (dashed line).}
\label{g3}
\end{center}
\end{figure}

We are now ready to derive the main results on the above expansion.
As in Sect.\ref{intro}, we fix the attention on the reparametrized  function
$N(\alpha, \delta,x)$ of Eq.\rref{repar} for
$\alpha \in (0,+\infty),~~\delta \in (-1, +\infty),~~x \in [0,+\infty)$
(the relations of $N$ with the standard Kummer function $M$ have been exploited in the
Introduction). \parn
We first discuss the case $\delta \geq 0$, where the function $\rho_n(\delta,\sq)$
of the previous Section is bounded both from below and above for all $n$.
\parn
\begin{prop}
\label{prmag}
\textbf{Proposition.} Let $\delta \in [0,+\infty)$, $\alpha \in (0,+\infty)$ and
$n \in \naturali_0$. For all $x \in [0,+\infty)$, it is
\beq  g_n(\alpha, \delta, x) \leq N(\alpha, \delta,x) \leq G_n(\alpha, \delta,x)
\label{demag1}~, \feq
where
\beq g_n(\alpha, \delta, x) := \sum_{k=0}^{n-1} \co{-\delta}{k} {\gamma(\alpha+k,x) \over x^{\alpha+k}} +
r_n(\delta) {\gamma(\alpha+n,x) \over x^{\alpha+n}}~,
\label{demag2} \feq
$$ G_n(\alpha, \delta, x) := \sum_{k=0}^{n-1} \co{-\delta}{k} {\gamma(\alpha+k,x) \over x^{\alpha+k}} +
R_n(\delta) {\gamma(\alpha+n,x) \over x^{\alpha+n}} $$
and $r_n$, $R_n$ are as in (\ref{rn}-\ref{rrnn}).
For $x=0$, each term $\gamma(\alpha+h,x)/x^{\alpha+h}$ in the above is intended to mean
its $x \vain 0^{+}$ limit, equal to $1/(\alpha+h)$.
\end{prop}
\textbf{Proof.} As already noted, $N(\alpha,\delta,x)$ is a Watson type integral
\rref{int}, with $b=1$ and
$\psi(t) \equiv \psi(\alpha,\delta;t) := t^{\alpha-1} (1-t)^{\delta}$.
On the other hand, Corollary \ref{coro} gives the inequalities
\beq
\sum_{k=0}^{n-1} {(-\delta)_{k}\over k!}~ t^k + r_n(\delta)~ t^{n}
\leq (1 - t)^{\delta}  \leq \sum_{k=0}^{n-1} {(-\delta)_{k}\over k!}~ t^k + R_n(\delta)~ t^{n}~,
\feq
whence
\beq \sum_{k=0}^{n-1} \co{-\delta}{k} t^{\alpha-1 + k} + r_n(\delta)~t^{\alpha-1 + n}
\leq t^{\alpha -1} (1 - t)^{\delta} \leq
\sum_{k=0}^{n-1} \co{-\delta}{k} t^{\alpha -1 + k} + R_n(\delta)~t^{\alpha -1 + n}~; \feq
thus, the thesis follows immediately from Prop.\ref{wat}.
\fine
We pass to the case $\delta < 0$, where the function $\rho_n(\delta, \sq)$ of the previous Section
is bounded only from below; in this case, a different method will be used to get
an upper bound on $N(\alpha, \delta,x)$.
\begin{prop}
\label{prmin}
\textbf{Proposition.} Let $\delta \in (-1,0)$, $\alpha \in (0,+\infty)$
and $n \in \naturali_0$. For all $x \in [0,+\infty)$, it is
\beq  g_n(\alpha, \delta, x) \leq N(\alpha, \delta,x) \leq G_n(\alpha, \delta,x)
\label{demin1}~. \feq
Here $g_n$ is as in Eq. \rref{demag2}, with $r_n(\delta) = (-\delta)_{n}/n!$ as prescribed  by \rref{agree};
furthermore
\beq G_n(\alpha, \delta, x) := \sum_{k=0}^{n-1} \co{-\delta}{k} {\gamma(\alpha+k,x) \over x^{\alpha + k}} +
S_n(\alpha, \delta) {\gamma(\alpha + n, x) \over x^{\alpha + n}}~, \label{demin2} \feq
\beq S_n(\alpha, \delta) := {(-\delta)_{n-1} \over (n-1)!}~
\left( {\alpha + n \over \delta + 1} - {\alpha + \delta + 1 \over n} \right)~.
\label{defs} \feq
(As usually, for $x=0$ one intends the previous definitions in a limit sense).
\end{prop}
\textbf{Proof.} In this case Corollary \ref{coro} implies
\beq \sum_{k=0}^{n-1} \co{-\delta}{k} t^{\alpha-1 + k} + r_n(\delta)~t^{\alpha-1 + n}
\leq t^{\alpha -1} (1 - t)^{\delta}~. \feq
Using again Prop.\ref{wat} on Watson integrals
we infer from here the lower bound $g_n$, of the form \rref{demag2}. \parn
In order to obtain the upper bound, we use a known recurrence relation for the
Kummer function. This is the identity $M(\alpha,\beta, x) = M(\alpha,\beta+1,x) + (\alpha x /\beta(\beta+1))
M(\alpha+1,\beta+2,x)$ \cite{Abr} that becomes, in terms of the reparametrized
function $N$,
\beq N(\alpha,\delta,x) = {\alpha + \delta + 1 \over \delta + 1}
N(\alpha, \delta + 1, x) - {x \over \delta + 1} N(\alpha + 1,\delta + 1,x)~.
\label{iden} \feq
The identity holds for all $\delta \in (-1,+\infty)$ and $\alpha \in (0,+\infty)$;
with the assumption $\delta \in (-1,0)$, the $N$ functions in the r.h.s.
depend on the parameter $\delta + 1 \in (0,1)$ and their expansions in terms
of Gamma functions can be performed via Prop.\ref{prmag}. More precisely,
we have
$$ N(\alpha, \delta + 1, x) \leq
\sum_{k=0}^{n-1} \co{-\delta -1}{k} {\gamma(\alpha+k,x) \over x^{\alpha+k}} +
R_n(\delta + 1) {\gamma(\alpha+n,x) \over x^{\alpha +n}}~, $$
\beq N(\alpha+1, \delta + 1, x) \geq
\sum_{k=0}^{n-1} \co{-\delta -1}{k} {\gamma(\alpha+k+1,x) \over x^{\alpha + k+1}} +
r_n(\delta + 1) {\gamma(\alpha+n+1,x) \over x^{\alpha + n +1}}~, \feq
$$ R_n(\delta +1) =
\co{-\delta -1}{n}~, \qquad r_n(\delta +1) =
- \cp{-\delta}{n-1}~. $$
Inserting these inequalities into the identity \rref{iden}, we get an upper
bound for $N(\alpha, \delta,x)$. This upper bound can be written as
in Eq.s (\ref{demin2}-\ref{defs}) reexpressing $\gamma(\alpha+k+1,x)$
and $\gamma(\alpha+n+1,x)$ by means of the recursion rule \rref{regam},
and then manipulating some of the occurring Pochhammer's symbols
via the identities \rref{repoc}.
\fine
\textbf{Remarks.} i) For $\delta \in (-1,0)$ and all $\alpha \in (0,+\infty)$,
it is $0 < r_n(\delta) < S_n(\alpha, \delta)$. \parn
ii) The remaining part of the Section will be devoted to the $n \vain +\infty$
limit in Prop.s \ref{prmag}, \ref{prmin}, yielding the convergent series expansion
\rref{series} for $N$. \parn
iii) Comparing the definitions \rref{repar}, \rref{beta} for $N$ and $B$, we see that
\beq N(\alpha,\delta,0)= B(\alpha,\delta+1)~. \feq
Thus, Prop.s \ref{prmag}, \ref{prmin} imply
\beq \sum_{k=0}^{n-1} \co{-\delta}{k} {1 \over \alpha + k} +
{r_n(\delta) \over \alpha + n} \leq B(\alpha,\delta+1) \leq
\sum_{k=0}^{n-1} \co{-\delta}{k} {1 \over \alpha + k} +
{R_n(\delta)~\mbox{or}~ S_n(\alpha,\delta) \over \alpha + n}~. \feq
For $n \vain +\infty$, this gives a series expansion for the Beta function.
\fine
\textbf{The error $\boma{\epsilon_n}$ in the expansion of $\boma{N}$.} For conveniency, let us define
\beq \epsilon_n(\alpha, \delta,x) := N(\alpha, \delta,x) -
\sum_{k=0}^{n-1} \co{-\delta}{k} {\gamma(\alpha+k,x) \over x^{\alpha+k}}~; \label{defep} \feq
this is the error in the approximation of $N$ by means
of $n$ incomplete Gammas. Prop.s \ref{prmag} and \ref{prmin} can be rephrased as
\beq e_n(\alpha, \delta, x) \leq \epsilon_n(\alpha, \delta,x) \leq E_n(\alpha, \delta, x)
\feq
for $\alpha \in (0,+\infty)$, $\delta \in (-1,+\infty)$, $x \in [0,+\infty)$,
$n \in \naturali_0$, where
\beq e_n(\alpha, \delta, x) := r_n(\delta) {\gamma(\alpha+n,x) \over x^{\alpha + n}}~, \qquad
E_n(\alpha, \delta, x) :=
\Big( R_n(\delta)~\mbox{or}~S_n(\alpha,\delta) \Big)~ {\gamma(\alpha+n,x) \over x^{\alpha + n}}~.
\label{errormag} \feq
The coefficient in the definition of $E_n$ is $R_n(\delta)$ for $\delta \in [0,+\infty)$, and
$S_n(\alpha, \delta)$ for $\delta \in (-1,0)$; as usually, $\gamma(\nu,x)/x^\nu := 1/\nu$
for $x=0$. The
quantities $e_n$, $E_n$ are the theoretical bounds
on the error $\epsilon_n$, derived from the previous Propositions.
Of course, by the asymptotics \rref{asi} of $\gamma$, fixing $n$ we have
\beq e_n(\alpha, \delta, x)  \sim r_n(\delta) {\Gamma(\alpha+n) \over x^{\alpha + n}},
E_n(\alpha, \delta, x)
\sim \Big( R_n(\delta)~\mbox{or}~S_n(\alpha,\delta) \Big)~ {\Gamma(\alpha+n) \over x^{\alpha + n}}
~~\mbox{for $x \vain + \infty$} \feq
(to get a better insight into $e_n$ and $E_n$ as functions of $x$, one could employ for $\gamma(\alpha+n,x)$
the approximants described in Sect.\ref{intro}, for example the Taylor
approximants for small $x$ and the Pad\'e ones for large $x$).
\parn
Let us discuss the error bounds when $n$ is large, starting from a
pointwise analysis in $x$ (a uniform analysis is be performed in the sequel).
\parn
\begin{prop}
\label{prerr}
\textbf{Proposition.} Fix $\alpha \in (0,+\infty)$, $\delta$ as below and $x \in [0,+\infty)$.
Then
\beq e_n(\alpha, \delta, x)
\left\{ \barray{lll} \sim e^{-x} /\delta \Gamma(-\delta) n^{1 + \delta}
&& \mbox{for $\delta \in (-1,+\infty)
\setminus \naturali$, $[\delta]$ even, $n \vain +\infty$,} \\
\sim e^{-x} / \Gamma(-\delta) n^{2 + \delta}
&& \mbox{for $\delta \in (-1,+\infty)
\setminus \naturali$, $[\delta]$ odd, $n \vain +\infty$,} \\
= 0 && \mbox{for $\delta \in \naturali$, $n \geq \delta + 1$;} \farray
\right. \feq
\beq E_n(\alpha, \delta, x)
\left\{ \barray{lll} \sim e^{-x} /\Gamma(-\delta) n^{2 + \delta}
&& \mbox{for $\delta \in (0,+\infty)
\setminus \naturali$, $[\delta]$ even, $n \vain +\infty$,} \\
\sim e^{-x} /\delta \Gamma(-\delta) n^{1 + \delta}
&& \mbox{for $\delta \in (0,+\infty)
\setminus \naturali$, $[\delta]$ odd, $n \vain +\infty$,} \\
\sim e^{-x} /(1+ \delta) \Gamma(-\delta) n^{1 + \delta}
&& \mbox{for $\delta \in (-1,0)$, $n \vain +\infty$,} \\
= 0 && \mbox{for $\delta \in \naturali$, $n \geq \delta + 1$.}
\farray \right. \feq
In any case $e_n(\alpha, \delta, x)$, $E_n(\alpha, \delta, x) \vain 0$, and thus
\beq \sum_{k=0}^{n-1} \co{-\delta}{k} {\gamma(\alpha+k,x) \over x^{\alpha+k}}
\vain N(\alpha, \delta, x) \qquad \mbox{for $n \vain +\infty$} \feq
(in the special case $\delta \in \naturali$, the sum in
the l.h.s. equals $N(\alpha,\beta,x)$ if $n -1 \geq \delta$).
\end{prop}
\textbf{Proof.} The definitions of the coefficients $r_n$, $R_n$ and $S_n$,
with Eq.s (\ref{eqsign1}-\ref{eqsign2}) and \rref{asipoc} on
the signs and asymptotics of the Pochhammer symbols, yield the
conclusions
\beq r_n(\delta)
\left\{ \barray{lll} \sim - 1/\Gamma(1-\delta) n^{\delta}  =
1/\delta \Gamma(-\delta) n^{\delta}
&& \mbox{$\delta \in (-1,+\infty)
\setminus \naturali$, $[\delta]$ even, $n \vain +\infty$,} \\
\sim 1/ \Gamma(-\delta) n^{1 + \delta} && \mbox{$\delta \in (-1,+\infty)
\setminus \naturali$, $[\delta]$ odd, $n \vain +\infty$}, \\
= 0 && \mbox{$\delta \in \naturali$, $n \geq \delta + 1$;} \farray
\right. \feq
\beq R_n(\delta)
\left\{ \barray{lll} \sim 1/\Gamma(-\delta) n^{1 + \delta}
&& \mbox{$\delta \in (0,+\infty)
\setminus \naturali$, $[\delta]$ even, $n \vain +\infty$,} \\
\sim - 1/ \Gamma(1-\delta) n^{\delta}
= 1/\delta \Gamma(-\delta) n^{\delta}
&& \mbox{$\delta \in (0,+\infty) \setminus \naturali$, $[\delta]$ odd, $n \vain +\infty$,} \\
= 0 && \mbox{$\delta \in \naturali$, $n \geq \delta + 1$;} \farray \right. \feq
\beq S_n(\alpha, \delta)
\sim {1 \over (\delta +1) \Gamma(-\delta) n^{\delta}},
\qquad \mbox{$\delta \in (-1,0)$, $n \vain +\infty$}. \feq
Furthermore, Eq.\rref{asial} implies
$\gamma(\alpha+n,x)/x^{\alpha+n} \sim e^{-x}/n$ for $n \vain +\infty$;
the thesis follows immediately.
\fine
Of course, the absolute value of the error $\epsilon_n$ can be bounded as well by means
of Prop.s \ref{prmag}, \ref{prmin}, which imply
\beq | \epsilon_n(\alpha, \delta, x) | \leq \EE_n(\alpha, \delta, x) := X_n(\alpha, \delta)
{\gamma(\alpha + n, x) \over x^{\alpha +n}}~, \label{epn} \feq
for all $\alpha \in (0,+\infty)$, $\delta \in (-1,+\infty)$, $x \in [0,+\infty)$,
where
\beq X_n(\alpha, \delta) := \left\{ \barray{lll}
\max ( {| (-\delta)_n | / n! }, {| (1 - \delta)_{n-1} | / (n-1)! } )
&& \mbox{for $\delta \in [0,+\infty)$}~, \\
S_n(\alpha, \delta) && \mbox{for $\delta \in (-1,0)$~.}~\farray \right.
\label{xn} \feq
The analogue of Prop.\ref{prerr} for the theoretical bound $\EE_n$ is the
following
\begin{prop}
\label{een}
\textbf{Proposition.} For fixed $\alpha \in (0,+\infty)$, $\delta$ as below and
$x \in [0,+\infty)$, it is
\beq \EE_n(\alpha, \delta, x)
\left\{ \barray{lll} \sim e^{-x}/| \delta \Gamma(-\delta) | n^{1 + \delta}
&& \mbox{for $\delta \in (0,+\infty) \setminus \naturali$, $n \vain +\infty$,} \\
\sim e^{-x} /(1+ \delta) \Gamma(-\delta) n^{1 + \delta}
&& \mbox{for $\delta \in (-1,0)$, $n \vain +\infty$,} \\
= 0 && \mbox{for $\delta \in \naturali$, $n \geq \delta + 1$.} \farray \right. \feq
\end{prop}
\textbf{Proof.} The thesis follows from the known behaviour of
$\gamma(\alpha+n,x)/x^{\alpha+n}$ and from the asymptotics
\beq X_n(\alpha, \delta)
\left\{ \barray{lll} \sim 1/| \delta \Gamma(-\delta) | n^{\delta}
&& \mbox{for $\delta \in (0,+\infty) \setminus \naturali$, $n \vain +\infty$,} \\
\sim 1 /(1+ \delta) \Gamma(-\delta) n^{\delta}
&& \mbox{for $\delta \in (-1,0)$, $n \vain +\infty$,} \\
= 0 && \mbox{for $\delta \in \naturali$, $n \geq \delta + 1$.} \farray \right. \label{asixn} \feq
\fine
\textbf{Reliability of the error bound $\boma{| \epsilon_n | \leq \EE_n}$: numerical tests.}
Of course, one would like the actual error $| \epsilon_n |$ to be close to
its theoretical estimator $\EE_n$. Figures \ref{g1}-\ref{g3}
report the ratio $| \epsilon_n |/\EE_n$
for fixed values of $\alpha, \delta, x$ as a function of $n$. They were generated
by the MATHEMATICA package, using its internal routines for computing numerically
the Kummer function and the incomplete Gammas. For reasons related to numerical accuracy, the
analysis has been limited to $1 \leq n \leq 13$; the outcomes seem to indicate that
$| \epsilon_n |/\EE_n$ approaches for large $n$ a constant value not far from $1$,
which was just the hoped result. For a better appreciation of the trends,
the points in the pictures corresponding to $n=1,2,...,13$ have been interpolated
by continuous or dashed lines, according to the considered values of $x$.
\vfill \eject \noindent
\textbf{Uniform analysis of the error.} We consider again
the error $\epsilon_n$ defined by \rref{defep}, and analyse it uniformly in $x$; to this purpose,
it is convenient to introduce an appropriate functional setting. For
$\sigma \in [0,+\infty)$, we consider the space
\beq \CC_{\sigma} := \{ f \in C([0,+\infty), \reali)~|~
+ \infty > \sup_{x \in (0, +\infty)} x^{\sigma} | f(x) | := \| f \|_{\sigma}~\}~; \feq
this is a Banach space with $\|~\|_{\sigma}$ as a norm.
We note that
\beq {\gamma(\nu, \sq) \over \sq^{\nu}}
\in \CC_{\sigma}~~\mbox{for $\nu >0$, $\sigma \in [0, \nu]$}~,\qquad
N(\alpha,\delta,\sq) \in C_{\sigma}~~\mbox{for
$\alpha >0, \delta >-1, \sigma \in [0,\alpha]$}
\feq
(in the case of $\gamma(\nu, \sq)/\sq^{\nu}$, this follows from
the asymptotics \rref{asi}; in the case of $N(\alpha,\delta,\sq)$
the statement follows from Prop.s \ref{prmag}, \ref{prmin} with
$n=1$, and from the asymptotics of the Gammas therein). \parn
The following result will be used to bind the $\sigma$-th norm of
$\gamma(\nu, \sq) / \sq^\nu$~.
\begin{prop}
\textbf{Lemma.} For all $\sigma,x \in [0,+\infty)$ and
$\mu \in (0,+\infty)$, it is
\beq 0 \leq {\gamma(\mu+\sigma,x) \over x^\mu} \leq
{\sigma^{\sigma} e^{-\sigma} \over \mu}
\feq
(intending this in a limit sense for vanishing $\sigma$ or $x$:
in particular, $\sigma^{\sigma} := 1$ if $\sigma=0$).
\end{prop}
\textbf{Proof.} The lower bound is obvious, let us prove the upper bound.
We can assume $\sigma>0$, and then recover the zero case taking the limit.
So, let $\sigma >0$;
due to \rref{asi}, the function $x \mapsto \gamma(\mu+\sigma,x)/x^\mu$
vanishes for $x$ approaching both $0^{+}$ and  $+\infty$,
and thus it attains its absolute maximum at a
point $c \in (0,+\infty)$; $c$ is a stationary point, i.e., a solution
of the equation
\beq 0 = {d \over d x} {\gamma(\mu+\sigma, x) \over x^\mu} =
{1 \over x} \left( x^{\sigma} e^{-x} - \mu
{\gamma(\mu+\sigma, x) \over x^\mu} \right)~. \label{stat} \feq
Thus, the maximum of the function is
\beq {\gamma(\mu +\sigma,c) \over  c^\mu} =
{c^{\sigma} e^{-c} \over \mu} \leq {\sigma^{\sigma} e^{-\sigma} \over \mu}~; \feq
the first equality follows from Eq.\rref{stat} and the subsequent inequality is elementary,
since the function $x \in [0,+\infty) \mapsto x^{\sigma} e^{-x}$ attains its maximum at
$x = \sigma$.
\fine
\textbf{Remark.} There is numerical evidence that the maximum point
$c = c_{\mu \sigma}$ of the above function is unique,
and that $\gamma(\mu+\sigma, c_{\mu \sigma})/(c_{\mu \sigma})^\mu$ is
not only bounded by $\sigma^{\sigma} e^{-\sigma}/ \mu$, but
in fact asymptotic to the latter for $\mu \vain +\infty$. \fine
\begin{prop}
\textbf{Corollary.} For all $\sigma \in [0,+\infty)$ and
$\nu \in (\sigma, +\infty)$, it is
\beq \left\| {\gamma(\nu, \sq) \over \sq^{\nu}} \right\|_{\sigma}
\leq {\sigma^{\sigma} e^{-\sigma} \over \nu - \sigma}~.\feq
\end{prop}
\textbf{Proof.} The above norm equals $\sup_{x \in [0,+\infty)} \gamma(\nu,x) /x^{\nu - \sigma}$;
now apply the previous Lemma. \fine
\begin{prop}
\textbf{Proposition.} Let $\alpha \in (0,+\infty)$,
$\delta \in (-1,+\infty)$, $\sigma \in [0,\alpha]$.
For all $n \in \naturali_0$ it is
\beq \left\| \epsilon_n(\alpha, \delta, \sq) \right\|_{\sigma}
\leq X_{n}(\alpha, \delta) {\sigma^{\sigma} e^{-\sigma} \over \alpha - \sigma + n}~,
\label{nob} \feq
with $X_n$ is as in Eq.\rref{xn}.
\end{prop}
\textbf{Proof.} Eq.\rref{epn} implies
\beq \left\| \epsilon_n(\alpha, \delta, \sq) \right\|_{\sigma}
\leq X_{n}(\alpha, \delta) \left\| {\gamma(\alpha + n, \sq) \over \sq^{\alpha+n}} \right\|_{\sigma}~;
\label{nobnor} \feq
now, use the previous Corollary.
\fine
\begin{prop}
\textbf{Proposition.} Let $\alpha \in (0,+\infty)$, $\delta \in (-1,+\infty)$,
$\sigma \in [0,\alpha]$. Then
\beq \mbox{r.h.s. of \rref{nob} }
\left\{ \barray{lll}
\sim \sigma^{\sigma} e^{-\sigma} / | \delta \Gamma(-\delta) | n^{1+\delta}
\vain 0 && \mbox{for $\delta \in (0,+\infty) \setminus \naturali$ and $n \vain +\infty$} \\
\sim \sigma^{\sigma} e^{-\sigma} / (1 + \delta) \Gamma(-\delta) n^{1+\delta}
\vain 0 && \mbox{for $\delta \in (-1,0)$ and $n \vain +\infty$} \\
= 0 && \mbox{for $\delta \in \naturali$ and $n \geq \delta +1$;}
\farray \right.
\label{nobas} \feq
thus
\beq  \sum_{k=0}^{n-1} \co{-\delta}{k} {\gamma(\alpha+k,\sq) \over \sq^{\alpha+k}}
\vain N(\alpha, \delta, \sq) \quad \mbox{for $n \vain +\infty$,~in the Banach space $\CC_{\sigma}$}. \feq
\end{prop}
\textbf{Proof.} Use the asymptotics \rref{asixn} for $X_n$. \fine
The accuracy of the expansion of $N$ via $n$ incomplete Gammas can be discussed
from a slightly different viewpoint, where the attention passes from the "absolute error"
$\epsilon_n$ to the relative uncertainty
\beq \xi_n(\alpha, \delta, x) := {G_n(\alpha, \delta, x) - g_n(\alpha, \delta, x) \over
| G_n(\alpha, \delta, x) + g_n(\alpha, \delta,x) |}~; \feq
here $G_n$ and $g_n$ denote, as usually, the upper and lower bounds for $N$ in
Prop.s \ref{prmag}, \ref{prmin} ({\footnote{A comment is worthy on the absolute value in the denominator
of $\xi_n$. The function $N$ is positive for all $\alpha, \delta$ and $x \in [0,+\infty)$,
so its upper bound $G_n$ is also positive. The lower bound $g_n$ and the sum
$g_n + G_n$ can be negative in some peculiar cases with large $\alpha, \delta$ and very small $n,x$.
For example, $G_1(10,10,0) + g_1(10,10,0) < 0$.}}).
From the expressions of these bounds, it is clear that
$\xi_n(\alpha, \delta, x) = O(1/x^n)$ for $x \vain +\infty$. \parn
In the next section, the $x$-dependence of $\xi_n$ will be appreciated numerically in a number of
examples.
\section{Low order approximants for $\boma{N}$.}
\label{exa}
\begin{figure}
\begin{center}
\includegraphics[
height=2.0in,
width=2.8in
]%
{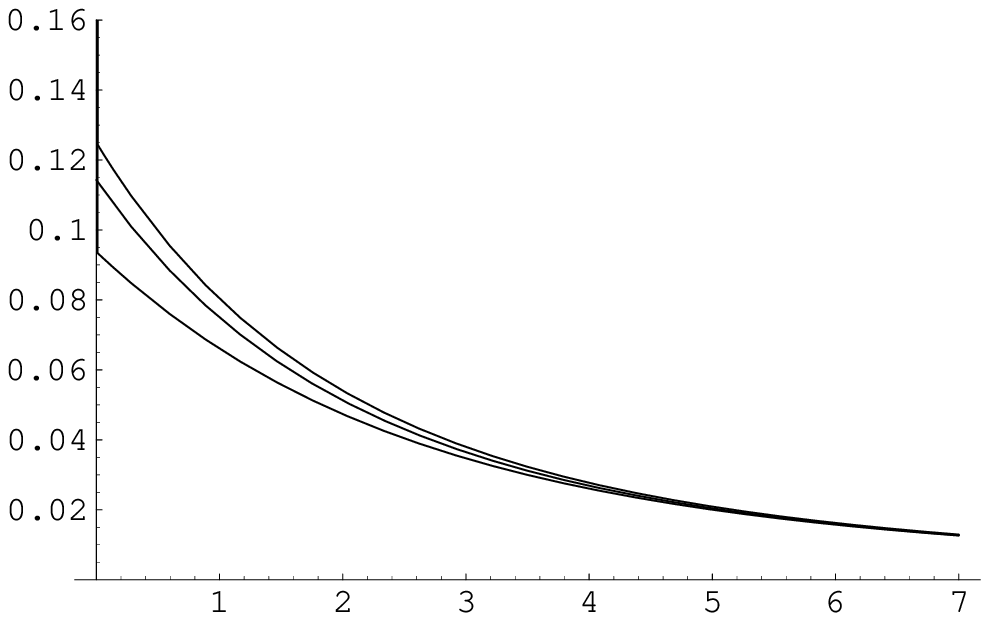}%
\caption{Graphs of $g(x)$, $N(2,3/2,x)$ and $G(x)$ from Example i), for $x \in [0,7]$.}
\label{f4}
\end{center}
\begin{center}
\includegraphics[
height=2.0in,
width=2.8in
]%
{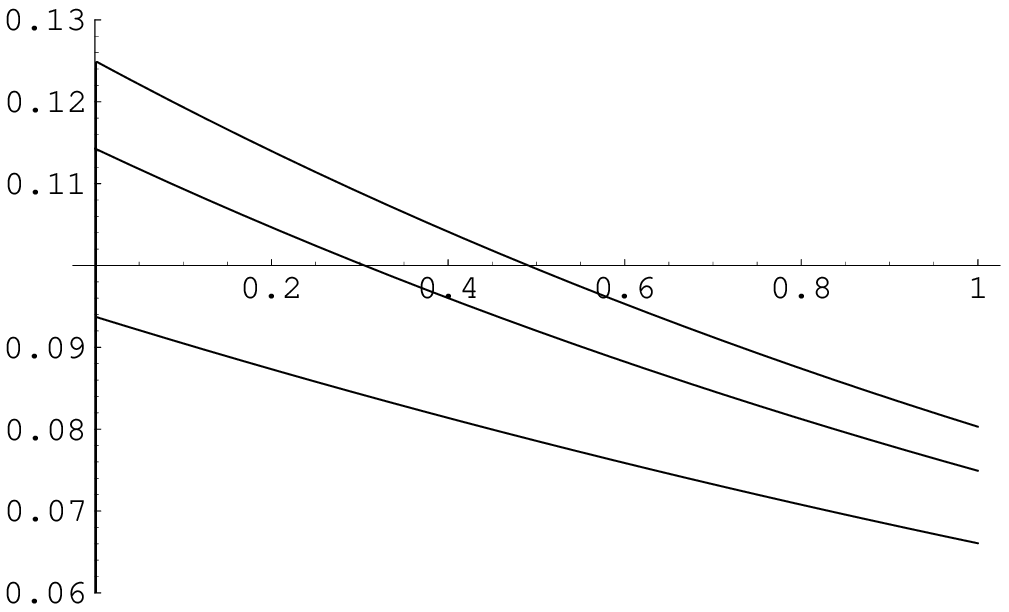}%
\caption{Graphs of $g(x)$, $N(2,3/2,x)$ and $G(x)$ from Example i), for $x \in [0,1]$.}
\label{f5}
\end{center}
\begin{center}
\includegraphics[
height=2.0in,
width=2.8in
]%
{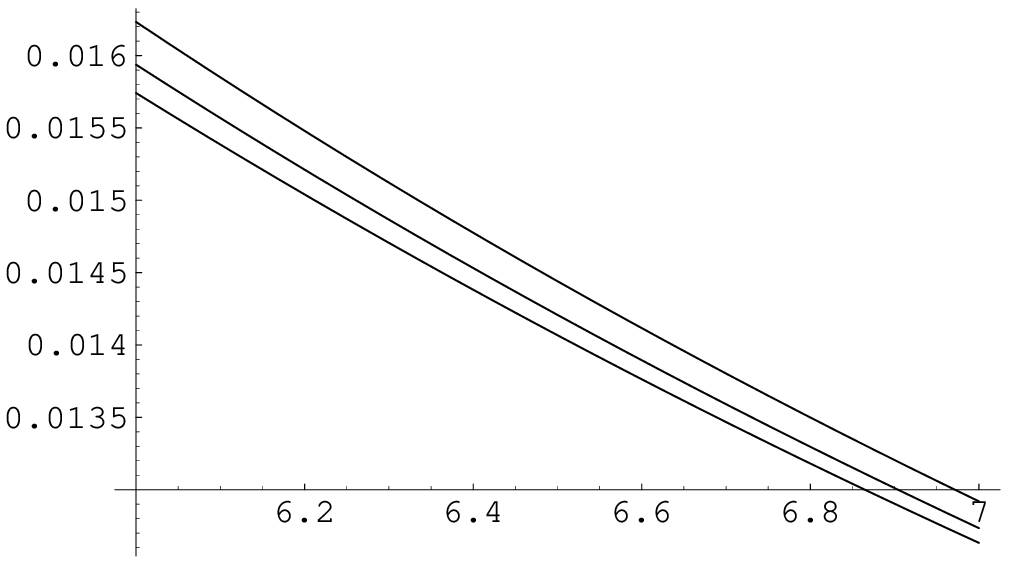}%
\caption{Graphs of $g(x)$, $N(2,3/2,x)$ and $G(x)$ from Example i), for $x \in [6,7]$.}
\label{f6}
\end{center}
\end{figure}

Let us show by some examples that Prop.s \ref{prmag}, \ref{prmin} give good approximants for $N$
even if they are used with low $n$.
The second example will show that a good accuracy
is kept even if we replace the incomplete Gammas with the approximants in terms of
exponentials, real powers of $x$ and rational functions reviewed in Sect.\ref{intro}
(in the other examples the Gammas are elementary functions, with exact expressions in
terms of exponentials and powers of $x$). In all the examples,
the low order approximants for $N$ will be compared with the Taylor ones (also presented
in Sect.\ref{intro}), which are found to be more precise for small $x$, namely,
for $x$ ranging from $0$ to some value of the order of unity.
\vskip 0.2cm \noindent
\textbf{i)} We consider the case $\alpha=2$, $\delta = 3/2$ and apply
Prop.\ref{prmag} with $n=2$. This gives
\beq g(x) \leq N(2,{3 \over 2},x) \leq G(x) \qquad \mbox{for all
$x \in [0,+\infty)$}~, \label{iaA}\feq
where
\beq g(x) := {\gamma(2,x) \over x^2} - {3 \gamma(3,x) \over 2 x^3} +
{3 \gamma(4, x) \over 8 x^4},~~
G(x) := {\gamma(2,x) \over x^2} - {3 \gamma(3,x) \over 2 x^3} +
{\gamma(4, x) \over 2 x^4} \feq
$$ \left(~ g(0) =  {3 \over 32}, \quad G(0) = {1 \over 8}~\right). $$
All the above incomplete Gammas are of integer order, so they can be computed by the
recursion method recalled in Sect.\ref{intro}:
$\gamma(2,x) = 1 - (x + 1)e^{-x}$, $\gamma(3, x) = 2 - (x^2 + 2 x +2) e^{-x}$,
$\gamma(4,x) = 6 - (x^3 + 3 x^2 + 6 x + 6) e^{-x}$. The conclusion is
\beq g(x) = {1 \over x^2} - {3 \over x^3} + {9 \over 4 x^4} +
\left({1 \over 8 x} + {7 \over 8 x^2} + {3 \over 4 x^3}
- {9 \over 4 x^4} \right) e^{-x}~, \feq
$$ G(x) = {1 \over x^2} - {3 \over x^3} + {3 \over x^4} +
\left({1 \over 2 x^2} - {3 \over x^4} \right) e^{-x}~. $$
Figure \ref{f4} contains the graphs of $g$, $G$ defined as above, and of
$N(2,3/2,\sq)$ as computed numerically by the MATHEMATICA package; the chosen interval for
$x$ is $[0,7]$. For a better
appreciation of the separation between $N(2,3/2,\sq)$ and the bounds $g$, $G$,
Figures \ref{f5} and \ref{f6}
report the graphs of the same functions for $x \in [0,1]$ and $x \in [6,7]$, respectively.
\parn
Let us describe the discrepancy of the bounds $g, G$ through the
relative uncertainty ({\footnote{here and in the forthcoming examples,
the bounds for which we evaluate the relative uncertainty are always positive,
for all $x \in [0,+\infty)$.}})
\beq \xi(x) := {G(x)-g(x) \over G(x) +g(x)}~. \feq
This function is decreasing: $\xi(0) = 1/7 < 0.15$,
$\xi(1) < 0.098$, $\xi(3) < 0.045$, $\xi(7) < 0.012$ and
$\xi(x) \vain 0$ for $x \vain +\infty$. \parn
The approximants for $N(2,3/2,x)$ constructed above can be improved for small $x$, using
in place of $g, G$ the Taylor bounds (\ref{tayn1}-\ref{tayn2}); with the choice $m=4$, these are
\beq t(x) \leq N(2, {3 \over 2}, x) \leq T(x) \qquad \mbox{for $x \in [0,+\infty)$}, \label{itT} \feq
\beq t(x) := {4 \over 35} - {16 \over 315} x + {16 \over 1155} x^2 - {128 \over 45045} x^3 +
{64 \over 135135} x^4 e^{-x}~, \feq
$$ T(x) :=
{4 \over 35} - {16 \over 315} x + {16 \over 1155} x^2 - {128 \over 45045} x^3 +
{64 \over 135135} x^4~. $$
The final bounds arising from \rref{iaA} and \rref{itT} are
\beq j(x) := \max(t(x),g(x)) \leq N(2,{3 \over 2},x) \leq J(x) :=
\min(T(x),G(x)) \feq
for all $x \in [0,+\infty)$. One finds numerically that: $j(x) = t(x)$ for $0 \leq x \leq 1.92$ and $j(x) = g(x)$
for $x \geq 1.93$; $J(x) = T(x)$ for $0 \leq x \leq 2.16$ and $J(x) = G(x)$
for $x \geq 2.17$. Let us introduce the relative uncertainty
\beq \eta(x) := {J(x)-j(x) \over J(x) + j(x)}~. \feq
Then: $\eta(0) = 0$, $\eta(1) < 0.0021$;
$\sup_{x \in [0,+\infty)} \eta(x) < 0.062$, the sup being attained at the point $x_0 \in (2.16, 2.17)$
where $T(x_0) = G(x_0)$; $\eta(x) = \xi(x)$ for $x \geq 2.17$.
\fine
\textbf{ii)} We keep $\delta = 3/2$ as before, but assume $\alpha=1/2$. Prop.\ref{prmag} with $n=2$ gives
\beq g(x) \leq N({1 \over 2},{3 \over 2},x) \leq G(x) \qquad \mbox{for all
$x \in [0,+\infty)$}~, \feq
where
\beq g(x) := {\gamma(1/2,x) \over \sqrt{x}} - {3 \gamma(3/2,x) \over 2 x^{3/2}} +
{3 \gamma(5/2, x) \over 8 x^{5/2}},
~~G(x) := {\gamma(1/2,x) \over \sqrt{x}} - {3 \gamma(3/2,x) \over 2 x^{3/2}} +
{\gamma(5/2, x) \over 2 x^{5/2}} \label{cota} \feq
$$ \left(~g(0) = {23 \over 20}, \quad G(0) = {6 \over 5}~\right). $$
One can express $\gamma(5/2,\sq)$, and $\gamma(3/2,\sq)$
in terms of $\gamma(1/2,\sq)$ by means of the recursion rule \rref{regam};
in this way, the previous bounds take the form
\beq g(x) = \left({1 \over \sqrt{x}}
- {3 \over 4 x^{3/2}} + {9 \over 32 x^{5/2}} \right) \gamma({1 \over 2},x) +
\left( {9 \over 8 x} - {9 \over 16 x^{2}} \right) e^{-x}, \label{copad} \feq
$$ G(x) =  \left({1 \over \sqrt{x}}
- {3 \over 4 x^{3/2}} + {3 \over 8 x^{5/2}} \right) \gamma({1 \over 2},x) +
\left( {1 \over x} - {3 \over 4 x^{2}} \right) e^{-x}~.
$$
From the above result we can infer upper and lower bounds for $N(1/2,3/2,x)$
in terms of elementary functions, using for $\gamma(1/2,x)$ the Pad\'e approximants
described in Sect.\ref{intro}. \parn
Let us recall the general Pad\'e approximants (\ref{pad1}-\ref{pad2}).
To get a lower bound for $g(x)$, we substitute in Eq.\rref{copad} the inequalities
$(1/\sqrt{x} + 9/32~ x^{-5/2}) \gamma(1/2,x) \geq$
$(1/\sqrt{x} + 9/32~ x^{-5/2}) \times$ (the Pad\'e
lower bound of order $q=1$),  $-3/4~ x^{-3/2} \gamma(1/2,x) \geq$
$-3/4~ x^{-3/2} \times$ (the Pad\'e upper bound of order $1$); we proceed
similarly to obtain an upper bound for $G(x)$, using again the approximants
with $q=1$. In this way, for all $x \in (0,+\infty)$ we find
$$ g(x)
\geq p(x) := {\sqrt{\pi} \over \sqrt{x}} - {3 \sqrt{\pi} \over 4 x^{3/2}}
+ {9 \sqrt{\pi} \over 32 x^{5/2}} +
\Big(~{9 \over 8 x} - {2 \over 3 + 2 x} +   $$
\beq + {3 \over 2 x (1 + 2 x)}
- {2 \over x (3 + 2 x)} - {9 \over 16 x^{2}} - {9 \over 16 x^{2} (3 + 2 x)}
- {9 \over 16 x^{3} (3 + 2 x)}\Big) e^{-x}~; \feq
$$ G(x)
\leq P(x) := {\sqrt{\pi} \over \sqrt{x}} - {3 \sqrt{\pi} \over 4 x^{3/2}}
+ {3 \sqrt{\pi} \over 8 x^{5/2}} +
\Big(~{1 \over x} - {2 \over 1 + 2 x} + $$
\beq - {3 \over 4 x^{2}} + {3 \over 2 x (3 + 2 x)}
- {3 \over 4 x^{2} (1 + 2 x)}
+ {3 \over 2 x^{2} (3 + 2 x)} \Big) e^{-x}~. \feq
The approximants $p, P$ are not accurate for small $x$ and tend to
$\mp \infty$, respectively, for $x \vain 0^{+}$. For this reason, we match them with the Taylor
bounds (\ref{tayn1}-\ref{tayn2}) of order $m=4$, which have the form
\beq t(x) \leq N({1\over 2},{3 \over 2},x) \leq T(x) \label{it2t2} \feq
\beq t(x) := {3 \pi \over 8} - {\pi \over 16} x + {3 \pi \over 256} x^2 - {\pi \over 512} x^3 +
{7 \pi \over 24576} x^4 e^{-x}~, \feq
$$ T(x) := {3 \pi \over 8} - {\pi \over 16} x + {3 \pi \over 256} x^2 - {\pi \over 512} x^3 +
{7 \pi \over 24576} x^4. $$
The final bounds for $N$ are
\beq \ell(x) := \max(t(x),p(x)) \leq N({1 \over 2},{3 \over 2},x) \leq L(x) :=
\min(T(x),P(x))~~\mbox{for $x \in [0,+\infty)$}. \feq
One finds numerically that: $\ell(x) = t(x)$ for $0 \leq x \leq 1.95$ and $\ell(x) = p(x)$
for $x \geq 1.96$; $L(x) = T(x)$ for $0 \leq x \leq 2.41$ and $L(x) = P(x)$
for $x \geq 2.42$. Let us introduce the relative uncertainty
\beq \vartheta(x) := {L(x)-\ell(x) \over L(x) + \ell(x)}~. \feq
Then: $\vartheta(0) = 0$, $\vartheta(1) < 0.00028$;
$\sup_{x \in [0,+\infty)} \vartheta(x) < 0.0074$, the sup being attained at the point $x_0 \in (2.41, 2.42)$
where $T(x_0) = P(x_0)$; $\vartheta(3) < 0.0051$, $\vartheta(7) < 0.0011$, $\vartheta(x) \vain 0$
for $x \vain +\infty$.
\fine
\textbf{iii)} We assume $\alpha=2$, $\delta = -1/2$ and apply
Prop.\ref{prmin}; in this case, to get satisfactory results it is
convenient to increase the order of the approximation, so we choose $n=4$.
In this way, we obtain
\beq g(x) \leq N(2,-{1 \over 2},x) \leq G(x) \qquad \mbox{for all
$x \in [0,+\infty)$}~, \label{idD}\feq
where
$$ g(x) := {\gamma(2,x) \over x^2} + {\gamma(3,x) \over 2 x^3} +
{3 \gamma(4, x) \over 8 x^4} + {5 \gamma(5, x) \over 16 x^5} + {35 \gamma(6, x) \over 128 x^6},~ $$
\beq G(x) :=  {\gamma(2,x) \over x^2} + {\gamma(3,x) \over 2 x^3} +
{3 \gamma(4, x) \over 8 x^4} + {5 \gamma(5, x) \over 16 x^5} + {455 \gamma(6, x) \over 128 x^6} \feq
$$ \left(~g(0) =  {667 \over 768}, \quad
G(0) = {1087 \over 768}~\right). $$
All the above incomplete Gammas are elementary: the ones of orders $2$, $3$, $4$ are read
from Example i), the other ones are $\gamma(5,x)= 24 - (x^4 + 4 x^3 + 12 x^2 + 24 x + 24) e^{-x}$,
$\gamma(6,x)= 120 - (x^5 + 5 x^4 + 20 x^3 + 60 x^2 + 120 x + 120) e^{-x}$.
Thus
\beq g(x) = {1 \over x^2} + {1 \over x^3} + {9 \over 4 x^4} + {15 \over 2 x^5} + {525 \over 16 x^6}
- \left({315 \over 128 x} + {735 \over 128 x^2} + {399 \over 32 x^3}
+ {837 \over 32 x^4} + {645 \over 16 x^5} + {525 \over 16 x^6} \right) e^{-x}, \feq
$$ G(x) = {1 \over x^2} + {1 \over x^3} + {9 \over 4 x^4} + {15 \over 2 x^5} + {6825 \over 16 x^6}
- \left({735 \over 128 x} + {2835 \over 128 x^2} + {2499 \over 32 x^3}
+ {7137 \over 32 x^4} + {6945 \over 16 x^5} + {6825 \over 16 x^6} \right) e^{-x}. $$
The relative difference
\beq \xi(x) := {G(x)-g(x) \over G(x) + g(x)}~\feq
is a decreasing function. It is: $\xi(0) = 210/877 < 0.24$, $\xi(1) <0.22$,
$\xi(3) < 0.15$, $\xi(7) < 0.046$ and $\xi(x) \vain 0$ for $x \vain +\infty$. \parn
Again, for small $x$ we can improve the previous estimates using the
Taylor bounds (\ref{tayn1}-\ref{tayn2}); with $m=4$, these are
\beq t(x) \leq N(2, -{1 \over 2}, x) \leq T(x) \qquad \mbox{for $x \in [0,+\infty)$}, \label{itdtd} \feq
\beq t(x) := {4 \over 3} - {16 \over 15} x + {16 \over 35} x^2 - {128 \over 945} x^3 +
{64 \over 2079} x^4 e^{-x}~, \feq
$$ T(x) := {4 \over 3} - {16 \over 15} x + {16 \over 35} x^2 - {128 \over 945} x^3 +
{64 \over 2079} x^4~. $$
The final bounds arising from \rref{idD} and \rref{itdtd} are
\beq j(x) := \max(t(x),g(x)) \leq N(2,- {1 \over 2},x) \leq J(x) :=
\min(T(x),G(x)) \feq
for all $x \in [0,+\infty)$.
One finds numerically that: $j(x) = t(x)$ for $0 \leq x \leq 1.57$ and $j(x) = g(x)$
for $x \geq 1.58$; $J(x) = T(x)$ for $0 \leq x \leq 1.54$ and $J(x) = G(x)$
for $x \geq 1.55$. The relative uncertainty
\beq \eta(x) := {J(x)-j(x) \over J(x) + j(x)} \feq
has the following features:
$\eta(0) = 0$, $\eta(1) < 0.016$;
$\sup_{x \in [0,+\infty)} \eta(x) < 0.20$, the sup being attained at the point $x_0 \in (1.57, 1.58)$
where $t(x_0) = g(x_0)$; $\eta(x) = \xi(x)$ for $x \geq 1.58$. \fine
\vfill \eject \noindent
\textbf{Acknowledgments.} We are grateful to S. Paveri-Fontana for useful discussions.
This work was partly supported by INdAM and by MIUR, COFIN/2001
Research Project "Geometry of Integrable Systems".

\end{document}